\documentclass[a4paper,12pt]{article}

\title{Some General Theorems of Incremental Thermoelectroelasticity}
\author{A. Montanaro}
\bigskip

\begin{document}  
\maketitle
Key words $\quad$    Thermoelectroelasticity,   Uniqueness of solution, Incremental thermoelectroelasticity, Hamilton principle,  Theorem of reciprocity of work.
\begin{abstract} 
We extend to incremental thermoelectroelasticity with biasing fields certain classical theorems, that have been stated and proved in linear thermopiezoelectricity referred to a natural configuration.
A uniqueness theorem for the solutions to the initial boundary value problem, the generalized Hamilton principle and a theorem of reciprocity of work are deduced for incremental fields superposed on finite biasing fields in a thermoelectroelastic body.
\end{abstract}
\section{Introduction}
In the last decades, 
with the increasing wide use in sensing and actuation, 
materials exhibiting couplings between elastic, electric, magnetic and thermal fields have attrached much attention.

In order to give certainty to experimental results and applications,
the interest of many researchers turned to the mathematical fitting of these topics.

Many applications have their mathematical formulation within a linear framework, and the theoretical study began from this context.

Foundamental is Nowacki's paper \cite{N:RTC}, where a uniqueness theorem for the solutions of the initial boundary value problems is proved in linear thermopiezoelectricity referred to a natural state, i.e., without biasing (or initial) fields. Hence Nowacki \cite{N:GTT}  also deduced a generalized Hamilton principle and a theorem of reciprocity of work. 

Li \cite{Li:URT} generalized the uniqueness and reciprocity theorems for linear thermo-electro-magneto-elasticity referred to a natural state.

Aouadi \cite{A:GTTM} establishes a reciprocal theorem for a linear theory in which the heat flux is considered as a constitutive independent variable, a rate-type evolution equation for it is added to the system of constitutive equations, and the entropy inequality is stated in the form proposed by M\"{u}ller \cite{M:CUF}. 

Iesan \cite{I:TED} uses the Green-Naghdi theory of thermomechanics of continua to derive a linear theory of thermoelasticity with internal structure where in particular a uniqueness result holds.

Related works on thermoelasticity and thermoelectromagnetism can be found in \cite{K:TPT} to \cite{MF:ECM}.

The classical linear theory of thermopiezoelectricity assumes infinitesimal deviations of the field variables from a reference state, where there are no initial mechanical and electric fields. In order to describe the response of thermoelectroelastic materials in the presence of initial fields one needs the theory for infinitesimal fields superposed on initial fields, and this can only be derived from the fully nonlinear theory of thermoelectroelasticity.
The equations of nonlinear thermoelectroelasticity were given in Tiersten \cite{T:NLETE}.
Yang \cite{Y:ESF} then derived from \cite{T:NLETE} the equations for infinitesimal incremental fields superposed on finite biasing fields in a thermoelectroelastic body with no assumption on the biasing fields.

Here we extend the aforementioned three Nowacki's theorems \cite{N:RTC}, \cite{N:GTT} to incremental 
thermoelectroelasticity with initial fields. 

We explicitly refer to the incremental theory \cite{Y:ESF}, hence we rewrite from this paper, with the same notations, some formulae and results on constitutive equations of incremental thermoelectroelasticity.

Of course, the theorems proved here just reduce to the ones in Nowacki's \cite{N:GTT} by neglecting the initial fields.

In the uniqueness theorem in Section \ref{section:unth} we assume that 
in the initial state entropy does not depend on time and temperature is uniform. For the theorem of reciprocity of work in Section \ref{section:RecWork} we assume that in the initial state both entropy and temperature fields do not depend on time.
\section{Equations of Nonlinear Thermoelectroelasticity}
\subsection{Balance laws and constitutive equations}
Consider a thermoelectroelastic body that, in the reference configuration, occupies a region $V$ with boundary surface $\,S$.
The motion of the body is described by $$y_i=y_i(X_L,\,t) \, ,$$
where $y_i$ denotes the present coordinates and $X_L$ the reference coordinates of material points with respect to the same Cartesian coordinate system.

Let $\;K_{Lj}, \; \rho_o, \; f_j,\; \Delta_L,\; \rho_E,\; \theta, \; \eta, \; Q_L\,$ and $\, \gamma\,$ respectively denote the first Piola-Kirchoff stress tensor, the mass density in the reference configuration, the body force per unit mass, the reference electric displacement vector, the free charge density per unit undeformed volume, the absolute temperature, the entropy per unit mass, the reference heat flux vector, and the body heat source per unit mass. 
Then we have the following equations of motion, electrostatics, and heat conduction written in material form with respect to the reference configuration:
\begin{equation}                    \label{eq:KLk}
	K_{Li,L}+\rho_o f_i=\rho_o\ddot y_i \, ,
\end{equation}
\begin{equation}                    \label{eq:DKk}
	\Delta_{L,L}=\rho_E \, ,
\end{equation}
\begin{equation}                    \label{eq:Q}
	\rho_o \theta \dot \eta =-Q_{L,L}+\rho_o\gamma \, ,
\end{equation}

The above equations are adjoined by constitutive relations defined by the specification of the free energy $\,\psi\,$ and heat flux $\,Q_L$:

\begin{equation}          \label{eq:psiEKL}
	\psi=\psi(E_{MN}, \, W_M,\, \theta) \, ,  \qquad 	Q_L=Q_L(E_{MN}, \, W_M,\, \theta, \, \Theta_M) \, 
\end{equation}
  where 
\begin{equation}    \label{eq:EKL}
E_{MN}=(y_{j,\,M}y_{j,\,N}-\delta_{MN})/2 \, , \qquad W_M=-\phi_M \, , 
\qquad \Theta_M=\theta_{,\,M}
\end{equation}
 are the finite strain tensor, the reference electric potential gradient, and the reference temperature gradient; of course,
 $\,\delta_{MN}\,$ is the Kronecker delta, and $\phi\,$ is the electric potential. 
 Hence, by using $\,\psi\,$ the constitutive relations $\,(4)\,$ of \cite{Y:ESF} are deduced  for 
 $\,K_{Li}, \; \Delta_{L}, \; \eta$; here we rewrite them from \cite{Y:ESF}:
 
 \begin{eqnarray}    \label{eqnarray:KLM}
K_{Li}=y_{i,\,A} \rho_o \frac{\partial \psi}{\partial E_{AL}}+JX_{L,\,j}\, \varepsilon_o(E_{j}E_{i}-\frac{1}{2}E_{i}E_{i}\delta_{ji}) \, ,  \nonumber \\
\Delta_{L}=\varepsilon_o JX_{L,\,j}E_{j} -\rho_o\frac{\partial \psi}{\partial W_{L}} \, ,   \\
\eta=-\frac{\partial \psi}{\partial \theta}  \, . \nonumber
\end{eqnarray}

Recall that the heat-flux constitutive relation  (\ref{eq:psiEKL})$_2$ is restricted by 
 
\begin{equation}    \label{eq:RQ}
Q_{L}\Theta_{L} \leq 0 \, .
\end{equation}

Note that, in particular,  (\ref{eq:psiEKL})$_2$ includes the case in which $\,Q_M\,$ is linear in 
$\,\Theta_{L} $, that is,
\begin{equation}          \label{eq:QTWG}
	Q_M=-\kappa_{ML}( \theta, \, W_A) \,\Theta_{L}  \, .
\end{equation}

\subsection{The initial boundary value problem for a thermoelectroelastic body}

To describe the corresponding boundary conditions to add to the field equations (\ref{eq:KLk})-(\ref{eq:Q}), three partitions 
$\,(S_{i1} ,  S_{i2})$, $\;i=1,\,2,\,3$, of the 
boundary surface $S=\partial {\mathcal B}$ can be assigned. 
 For mechanical  boundary conditions, deformation $\, \tilde y_i\,$   
 and traction 
 $\,\tilde t_i\,$  per unit undeformed area 
 are prescribed, respectively, on $\,S_{11}\,$ and  $\, S_{12}$;
  for electric  boundary conditions,  electric potential 
  $\,\tilde{\phi}\,$  and surface-free charge  
  $\,\tilde{\Delta} \,$   per unit undeformed area 
  are prescribed, respectively,  on $\,S_{21}\,$ and  $\, S_{22}$;
  while for thermic boundary conditions,  temperature  
   $\,\tilde{\theta}\,$
   and normal heat flux 
    $\,\tilde{Q}\,$ per unit undeformed area 
  are prescribed, respectively,on $\,S_{31}\,$ and  $\, S_{32}$.
  Hence, we can write
 \begin{equation}    \label{eq:yiyi}
y_i= \tilde y_i \quad {\rm on}  \quad S_{11} \, ,	\qquad K_{Li}N_{L}= \tilde K_i \quad {\rm on} \quad  S_{12}  \quad  \quad {\rm ('mechanical')}\, , 
\end{equation}
 \begin{equation}    \label{eq:phi}
	\phi=\tilde{\phi} \quad {\rm on}  \quad S_{21} \,	,\qquad 
	\Delta_{L}N_{L} =-\tilde{\Delta} \quad {\rm on}  \quad S_{22} \,,\qquad \quad {\rm ('electric')}	\\
\end{equation}
 \begin{equation}    \label{eq:T}
	\theta =\tilde{	\theta} \quad {\rm on}  \quad S_{31}\,	,\qquad 
Q_{L}N_{L}=\tilde{Q} \quad {\rm on}  \quad S_{32} \qquad \quad {\rm ('thermic')}\,,	\\
\end{equation}
 where $\,{\bf N}=(N_{L})\,$ is the unit exterior normal on $\,S\,$ and
 \begin{equation}    \label{eq:SyS}
S_{i1} \cup  S_{i2}
=S \,,\quad
S_{i1} \cap  S_{i2}=\emptyset \, \quad (i=1,\,2,\,3) \,.
		\end{equation} 
 
 We put 
\begin{equation}    \label{eq:exactionbd}
 {\mathcal A}_{body}:= \Big( \, f_i, \, \rho_E,\, \gamma \,\Big) \, ,
\end{equation}
\begin{equation}    \label{eq:exactionsf}
 {\mathcal A}_{surf}:= \Big( \, \tilde y_i, \, \tilde K_i,\, \tilde{\phi} , \, \tilde{\Delta} , \, \tilde{\theta},\, \tilde{Q}\,\Big)  \, ,
\end{equation}
\begin{equation}    \label{eq:exaction}
 {\mathcal A}:= \big(\,{\mathcal A}_{body}, \,  {\mathcal A}_{surf}\, \big)
 =\Big( \, f_i, \, \rho_E,\, \gamma, \, \tilde y_i, \, \tilde K_i,\, \tilde{\phi} , \, \tilde{\Delta} , \, \tilde{\theta},\, \tilde{Q}\,\Big)  \, .
\end{equation}
$\,{\mathcal A}_{body}$,  $\,{\mathcal A}_{body}$, and $\,{\mathcal A}\,$ are said to be the {\it (external) body-action}, {\it surface-action},
 and {\it action}, respectively.
 
 The initial conditions have the form
\begin{eqnarray}         \label{eqnarray:incond}  \nonumber
	y_i({\bf X}, \, 0)=f_i({\bf X}), \quad 	\dot y_i({\bf X}), \, 0)=g_i({\bf X}), \\ 
	\quad 	\theta({\bf X}, \, 0)=h({\bf X}), \quad 
		\quad 	\phi({\bf X}, \, 0)=l({\bf X}) \qquad ({\bf X} \in B, \; t=0)\,,
\end{eqnarray}
 where $${\mathcal I}=\Big( f_i,\ g_i,\, h,\, l\Big)$$
 are prescribed smooth functions of domain $\,V$.
The initial boundary value problem is then stated as:
{\it assigned $\,{\mathcal A}_{body}$, to find the solution $\,(\phi, \, \theta, \, y_i)\,$ in $\,{\mathcal B}\,$ to the constitutive relations (\ref{eqnarray:KLM}) and field equations  (\ref{eq:KLk})-(\ref{eq:Q}) which satisfies the boundary conditions  (\ref{eq:yiyi})-(\ref{eq:T}) and initial conditions 
(\ref{eqnarray:incond}) for given $\,{\mathcal A}_{surf}\,$ and $\,{\mathcal I}$.}

\section{Biasing and incremental fields}
In incremental theories three configurations are distinguished: the reference, initial and present configuration. 
\subsection{The Reference Configuration}
In the reference state the body is undeformed and free of all fields. A generic point at this state is denoted by 
{\bf X} with rectangular coordinates $X_N$.
The mass density in the reference configuration is denoted by $\,\rho_o$.

\subsection{The Initial Configuration}
 We put 
\begin{equation}    \label{eq:exactionbdo}
 {\mathcal A}^o_{body}:= \Big( \, f^o_i, \, \rho_E^{o},\, \gamma^o  \,\Big)   \, ,
\end{equation}
\begin{equation}    \label{eq:exactionsfo}
 {\mathcal A}^o_{surf}:= \Big( \, \tilde y^o_i, \, \tilde K^o_i,\, \tilde{\phi}^o , \, \tilde{\Delta}^o , \, \tilde{\theta}^o,\, \tilde{Q}^o\,\Big)  \, ,
\end{equation}
\begin{equation}    \label{eq:exactiono}
 {\mathcal A}^o:=  \big(\,{\mathcal A}_{body}, \,  {\mathcal A}_{surf}\, \big)
 =\Big( \, f_i^o, \, \rho_E^{o},\, \gamma^o, \, \tilde y_i^o, \, \tilde K_i^o,\, \tilde{\phi}^o , \, \tilde{\Delta}^o , \, \tilde{\theta}^o,\, \tilde{Q}^o\,\Big) \, .
\end{equation}
 $\,{\mathcal A}^o_{body}$,  $\,{\mathcal A}^o_{body}$, and $\,{\mathcal A}^o\,$ are said to be the {\it (external) body-action}, {\it surface-action},
 and {\it action}, respectively.

In this state the body is deformed finitely under the action of the prescribed action  $\,{\mathcal A}^o$.
The position of the material point associated with $\, {\bf X} \,$ is given by   $$\,y^o_\alpha = y^o_\alpha({\bf X}, \,t)\,, $$ 
with the Jacobian of the initial configuration denoted by $$\,J_o=det(y^o_{\alpha, \,L})\, . $$

In this state the electric potential, electric field and temperature field are denoted by
$\,\phi^o({\bf X}, \,t),\; W_\alpha^o=-\phi^o_{,\,\alpha}\,$ and $\,\theta^o({\bf X}, \,t)$, respectively.

The initial fields
\begin{equation}       \label{eq:infields}
 y^o_\alpha = y^o_\alpha({\bf X}, \,t), \quad \phi^o=\phi^o({\bf X}, \,t), \quad \theta^o=\theta^o({\bf X}, \,t)
 \end{equation}
satisfy the equations of nonlinear thermoelectroelasticity (\ref{eq:KLk})-(\ref{eq:SyS})  under the prescribed action $\,{\mathcal A}^o$.  
In studying the incremental fields the solution to the initial state problem is assumed known.
 
\subsection{The Present Configuration}
 To the deformed body at the initial configuration, infinitesimal deformations, electric, and thermal fields are applied. 
The present position of the material point associated with  $\, {\bf X} \,$ is given by $\,y_i({\bf X},\,t)$, with electric potential  $\,\phi({\bf X},\,t)\,$ and temperature $\,\theta({\bf X},\,t)$.

The fields  $\,y_i({\bf X},\,t), \; \phi({\bf X},\,t), \; \theta({\bf X},\,t)\,$ satisfy (\ref{eq:KLk})-(\ref{eq:Q}) under the action of the external action (\ref{eq:exaction}).

\subsection{Equations for the incremental fields}
 Let $\varepsilon$ be a small and dimensionless number. 
The incremental process $\,\varepsilon(y^1,\, \phi^1,\, \theta^1)\,$ for $\,(y,\, \phi,\, \theta)\,$  superposed to the initial process $\,(y^o,\, \phi^o,\, \theta^o)\,$  is assumed to be infinitesimal and, therefore, we write:
\begin{equation}    \label{eq:yidialp}
y_{i}= \delta_{i\alpha}(y^o_\alpha + \varepsilon y^1_\alpha) \, , \qquad 
\phi= \phi^o + \varepsilon \phi^1 \, ,
\qquad  \theta= \theta^o + \varepsilon \theta^1 \, ,
\end{equation}
Corresponding to (\ref{eq:yidialp}), the other quantities of the present state can be written as:

 \begin{equation}    \label{eq:acto1}  \nonumber
  {\mathcal A}\cong  {\mathcal A}^o + \varepsilon  {\mathcal A}^1\, ,\\
\end{equation}
where, due to nonlinearity, higher powers of $\varepsilon$ may arise.
For the incremental action we have 
\begin{equation}    \label{eq:exactionbd1}
 {\mathcal A}^1_{body}:= \Big( \, f^1_i, \, \rho_E^{1},\, \gamma^1  \,\Big)
\end{equation}
\begin{equation}    \label{eq:exactionsf1}
 {\mathcal A}^1_{surf}:= \Big( \, \tilde y^1_i, \, \tilde K^1_i,\, \tilde{\phi}^1 , \, \tilde{\Delta}^1 , \, \tilde{\theta}^1,\, \tilde{Q}^1\,\Big)
\end{equation}
\begin{equation}    \label{eq:exaction1}
 {\mathcal A}^1:=  \big(\,{\mathcal A}_{body}, \,  {\mathcal A}_{surf}\, \big)
 =\Big( \, f_i^1, \, \rho_E^{1},\, \gamma^1, \, \tilde y_i^1, \, \tilde K_i^1,\, \tilde{\phi}^1 , \, \tilde{\Delta}^1 , \, \tilde{\theta}^1,\, \tilde{Q}^1\,\Big)
\end{equation}
We want to derive equations governing the incremental process 

$$\,\big( {\bf u}:={\bf y}^1 \,, \quad \phi^1 \, , \quad \theta^1 \,\big) \,.$$

From (\ref{eq:yidialp}) and (\ref{eq:acto1}), we can further write:

 \begin{eqnarray}    \label{eqnarray:EKL}   \nonumber
E_{KL} \cong E^o_{KL} + \varepsilon E^1_{KL} \, ,  \\ 
W_{L} \cong W^o_{L} + \varepsilon W^1_{L} \, ,  \\ \nonumber
\Theta_{L} \cong \Theta^o_{L} + \varepsilon \Theta^1_{L} \, ,  \\ \nonumber
 \end{eqnarray}
where
 \begin{eqnarray}    \label{eqnarray:E0E1KL} \nonumber
E^o_{KL} =  ( y^o_{\alpha , \,K} y^o_{\alpha , \,L} - \delta_{KL})/2 \, , \qquad
E^1_{KL} =  ( y^o_{\alpha , \,K} y^1_{\alpha , \,L} + y^o_{\alpha , \,L} y^1_{\alpha , \,K})/2 \, , \\   
W^o_{L} =  -\phi^o_{,\, L}\, , \qquad W^1_{L} =  -\phi^1_{,\, L}                 \, , \\ \nonumber
\Theta^o_{L} =  \theta^o_{,\, L}\, , \qquad \Theta^1_{L} =  \theta^1_{,\, L}   \, . \\   \nonumber
 \end{eqnarray}

Substituting (\ref{eq:yidialp})-(\ref{eqnarray:E0E1KL}) into the constitutive relations (\ref{eq:KLk})-(\ref{eq:Q}),
with some very lengthy algebra, the following expression are obtained \cite{Y:ESF}:
 \begin{eqnarray}    \label{eqnarray:KKLdelt}    \nonumber
K_{Mi} \cong \delta_{i \alpha} (K^o_{M\alpha} + \varepsilon K^1_{M \alpha}) \, ,  \quad
\Delta_{M} \cong \Delta^o_{M} + \varepsilon \Delta^1_{M} \, , \\
\eta \cong \eta^o + \varepsilon \eta^1 \, ,  \quad
Q_{M} \cong Q^o_{M} + \varepsilon Q^1_{M} \, .  
 \end{eqnarray}
where
 \begin{equation}    \label{eq:K1KalpaG} 
K^1_{M \alpha} =  G_{M \alpha L \gamma} u_{\gamma, \, L} +  R_{LM \alpha} \phi^1_{, \, L} - \rho_o \Lambda_{M \alpha}\theta^1  \, ,
 \end{equation}
 \begin{equation}    \label{eq:D1RLP} 
\Delta^1_{M} =  R_{M N \gamma} u_{\gamma, \, N} - L_{MN} \phi^1_{, \, N} + \rho_o P_{M}\theta^1  \, ,
 \end{equation}
 \begin{equation}    \label{eq:Eta1P} 
\eta^1 =  \Lambda_{M \gamma} u_{\gamma, \, M} -P_{M} \phi^1_{, \, M} + \alpha \theta^1  \, ,
 \end{equation}
 \begin{equation}    \label{eq:Q1ABCF} 
Q^1_{M} =  A_{M N \alpha} u_{\alpha, \, N} - B_{MN} \phi^1_{, \, N} + C_M\theta^1 + F_{MN} \theta^1_{,\,N}  \, .
 \end{equation}
 
 By putting 
 $$\kappa_{MN\alpha}=-A_{M N \alpha}, \quad \kappa^E_{MN}=B_{MN},\quad
 \kappa_{M}=-C_M, \quad \kappa_{MN}=-F_{MN},$$  
 the latter rewrites as
    \begin{equation}    \label{eq:Qflow} 
Q^1_{M} =-\kappa_{MN\alpha}u_{\alpha, \,N} -\kappa^E_{MN} \phi^1_{, \, N} - \kappa_{M} \theta^1 - \kappa_{MN} \theta^1_{,\,N}  \, .
 \end{equation}
In (\ref{eq:K1KalpaG})-(\ref{eq:Q1ABCF}),
$\; G_{M \alpha L \gamma}\,$ are the effective elastic constants,
$\, R_{LM \alpha}\,$ are the effective piezoelectric constants,
$\, \Lambda_{M \alpha}\,$ are the effective thermoelatic constants,
$\,L_{MN}\,$ are the effective dielectric constants,
 $\, P_{M}\,$ are the effective pyrolectric constants,
 $\, \alpha\,$ is related with the specific heat.
Their expressions are \cite{Y:ESF}:
\begin{eqnarray}          \label{eqnarray:exprconstants}  \nonumber 
	G_{K \alpha L \gamma}\,=\,y^o_{\alpha,\,M}\rho_o
	           \frac{\partial^2 \psi}{\partial E_{KM} \partial E_{LN}}(\theta^o, \, E^o_{AB}, \, W^o_{A})\,y^o_{\alpha,\,L} 
	           +  
	  \rho_o \frac{\partial \psi}{\partial E_{KL}}(\theta^o, \, E^o_{AB}, \, W^o_{A})\,\delta_{\alpha \gamma} + g_{K \alpha L \gamma}
	           \,,           \qquad \qquad  \qquad \qquad     \qquad \qquad        \qquad \qquad   \qquad \qquad \qquad \qquad    \qquad      \nonumber  \\
	R_{LM \gamma}= -\rho_o \frac{\partial^2 \psi}{\partial W_{K} \partial E_{ML}}(\theta^o, \, E^o_{AB}, \, W^o_{A})\,y^o_{\gamma,\,M}+ r_{K L \gamma}  , \qquad  \qquad \qquad     \qquad \qquad     \qquad \qquad     \qquad \qquad     \qquad \qquad    \qquad \qquad  \qquad \qquad    \nonumber \\ 
\Lambda_{M \gamma} \,=\, -\frac{\partial^2 \psi}{\partial E_{LM} \partial \theta}(\theta^o, \, E^o_{AB}, \, W^o_{A})\,y^o_{\gamma,\,L} \, ,
 \qquad \qquad     \qquad \qquad     \qquad \qquad     \qquad \qquad \qquad \qquad     \qquad \qquad         \qquad \qquad      \nonumber  \\
	L_{MN}\,=\,  -\rho_o \frac{\partial^2 \psi}{\partial W_{M} \partial W_{N}}(\theta^o, \, E^o_{AB}, \, W^o_{A})+ l_{MN} \, ,  \quad
	P_{M} \,=\, -\frac{\partial^2 \psi}{\partial W_{M} \partial \theta}(\theta^o, \, E^o_{AB}, \, W^o_{A}) \, ,
	 \qquad \qquad     \qquad \qquad \qquad \qquad     \qquad \qquad     \qquad \qquad     \qquad \qquad     \qquad    \qquad    \qquad     \\
\alpha \,=\, -\frac{\partial^2 \psi}{\partial  \theta^2}(\theta^o, \, E^o_{AB}, \, W^o_{A}) \, ,\quad  A_{M N \gamma} \,=\, \frac{\partial Q_M}{\partial E_{LN}}(\theta^o, \, E^o_{AB}, \, W^o_{A})\, y^o_{\gamma, \, L}   \,=\, - \kappa_{MN\gamma}  \,  ,
    \qquad  \qquad \qquad     \qquad \qquad     \qquad \qquad     \qquad \qquad     \qquad \qquad    \qquad \qquad  \qquad \qquad    \nonumber \\ 
    B_{MN}\,=\, \frac{\partial Q_{M}}{\partial W_{N}}(\theta^o, \, E^o_{AB}, \,
 W^o_{A})\,=\,  \kappa^E_{MN}  \,  ,  
    \qquad  \qquad \qquad     \qquad \qquad     \qquad \qquad     \qquad \qquad     \qquad \qquad     \qquad \qquad  \qquad \qquad  \qquad \qquad   \nonumber \\   
C_{_{M}}\,=\, \frac{\partial Q_{M}}{\partial \theta}(\theta^o, \, E^o_{AB}, \, W^o_{A})\,=\, - \kappa_{M}  \,  ,  \qquad  \qquad 
  F_{MN}\,=\, \frac{\partial Q_{M}}{\partial \Theta_{N}}(\theta^o, \, E^o_{AB}, \, W^o_{A})\, =\, - \kappa_{MN}  \,  ,  \qquad  \qquad  \qquad  \qquad  \qquad  \qquad  \qquad  \qquad  \qquad  \qquad  \qquad  \qquad  \qquad \qquad   \nonumber  
\end{eqnarray}
where 
\begin{eqnarray} \label{eqnarray:gprp}  \nonumber 
	g_{K \alpha L \gamma}\,=\, 
	\varepsilon_o J_o \Big[W^o_\alpha W^o_\beta \big(X_{K,\,\beta}X_{L,\,\gamma}-X_{K,\,\gamma}X_{L,\,\beta}\big)  
	                      +W^o_\beta W^o_\gamma \big(X_{K,\,\alpha}X_{L,\,\beta}-X_{K,\,\beta}X_{L,\,\alpha} \big)\qquad \nonumber \\
	                      +W^o_\beta W^o_\beta \big(X_{K,\,\gamma}X_{L,\,\alpha}-X_{K,\,\alpha}X_{L,\,\gamma} \big)/2 
	                      -W^o_\alpha W^o_\gamma X_{K,\,\beta}X_{L,\,\beta} \Big] \, ,  \qquad  \qquad  \\
	                      	r_{K L \gamma}\,=\, 
	\varepsilon_o J_o \Big(W^o_\alpha X_{K,\,\alpha}X_{L,\,\gamma}- W^o_\alpha X_{K,\,\gamma}X_{L,\,\alpha} - W^o_\gamma X_{K,\,\alpha}X_{L,\,\alpha} \Big) \, , \quad
	l_{MN}\,=\,	\varepsilon_o J_o X_{M,\,\alpha}X_{N,\,\alpha}\,.   \nonumber \\ \nonumber
\end{eqnarray}
In (\ref{eq:Q1ABCF}) we have introduced the $\, \kappa$-notation to allow comparison between the proofs written here and those written in \cite{N:GTT}.
The following symmetries hold:
\begin{equation}    	\label{eq:symmetries} 
		G_{K \alpha L \gamma}\,=\,	G_{L \gamma K \alpha}\,,  \qquad 	L_{MN}\,=\,	L_{NM} \, .
\end{equation}
\subsection{Restriction on the incremental heat flux}
Now we show that the restriction (\ref{eq:RQ}) on the heat flux
(\ref{eq:psiEKL})$_2$, together with the condition 
\begin{equation}   \label{eq:Q=0}
Q^o_{L}=0 \quad {\rm for} \quad \Theta^o_{L}=0 \,,
\end{equation}
implies an analogous restriction on the incremental heat flux (\ref{eq:Q1ABCF}), that is,
\begin{equation}    \label{eq:RQ1}
Q^1_{L}\Theta^1_{L} \leq 0 \, .
\end{equation} 
Indeed, substituting 
$$Q_{L}=Q^o_{L}+ \varepsilon Q^1_{L}\,, \qquad \Theta_{L}=\Theta^o_{L}+ \varepsilon \Theta^1_{L}\,$$
in (\ref{eq:RQ}),  we obtain
\begin{equation}
	\Big(Q^o_{L}+ \varepsilon Q^1_{L}\Big)\Big(\Theta^o_{L}+ \varepsilon \Theta^1_{L}\Big) \leq 0  \, ,
\end{equation}
which for $\, \Theta^o_{L}=0$,   by (\ref{eq:Q=0}),  yields (\ref{eq:RQ1}).

Note that the choice (\ref{eq:QTWG}) for the heat flux response function
satisfies (\ref{eq:Q=0}).

\subsection{Incremental field equations}
By substituting (\ref{eq:yidialp})-(\ref{eqnarray:KKLdelt}) into (\ref{eq:KLk})-(\ref{eq:Q}) and (\ref{eq:yiyi})-(\ref{eq:T}),
we find the governing equations for the incremental fields 
\begin{equation}    \label{eq:k1}  
K^1_{M \alpha, \, M}\,+\, \rho_o \, f^1_{\alpha}\,=\,\rho_o \, \ddot u_{\alpha}\, ,
\end{equation}  
\begin{equation}    \label{eq:D1}  
\Delta^1_{M, \, M}\,=\, \rho^1_{E} \, ,
\end{equation} 
\begin{equation}    \label{eq:N1}  
\rho_o \,( \theta^o \dot \eta^1 \,+\, \theta^1 \dot \eta^o ) \,=\,-Q^1_{M, \, M} \,+\, \rho_o \, {\gamma}^1 \, .
\end{equation}  
Introducing the constitutive relations (\ref{eq:K1KalpaG})-(\ref{eq:Q1ABCF}) into the incremental equations of motion (\ref{eq:k1}), the equation of the electric field (\ref{eq:D1}), and the heat equation (\ref{eq:N1}), for  $\,f^1_\alpha=0\,$ we have
\begin{equation}    \label{eq:k111}  
G_{M \alpha L \gamma} u_{\gamma, \, LM} 
+  R_{LM \alpha} \phi^1_{, \, LM} 
- \rho_o \Lambda_{M \alpha}\theta^1_{, \,M} \,=\,\rho_o \, \ddot u_{\alpha}\, ,
\end{equation}  
\begin{equation}    \label{eq:D111}  
R_{M N \gamma} u_{\gamma, \, NM} - L_{MN} \phi^1_{, \, NM} + \rho_o P_{M}\theta^1_{M} \,=\, \rho^1_{E} \, ,
\end{equation}  
\begin{eqnarray}    \label{eqnarray:N111}  
\rho_o \theta^o \Big(\, \Lambda_{M \gamma} \dot u_{\gamma, \, M} -P_{M} \dot \phi^1_{, \, M} + \alpha \dot \theta^1\,\Big) \,+\,
\rho_o \theta^1 \dot \eta^o   \nonumber \\
 \,=\,\kappa^E_{MN} \phi^1_{, \, NM} +\kappa_M\theta^1_{, \,M} +
\kappa_{MN} \theta^1_{,\,NM} +
\kappa_{MN\alpha} u_{\alpha,\,NM} +\rho_o \, {\gamma}^1 \, . 
\end{eqnarray}  
\section{Uniqueness theorem of the solution of the incremental differential equations}  \label{section:unth} 
In the present Section we assume $\,\dot \eta^o =0\,$ and
$\,\Theta^o_L=0$, i.e. the initial temperature field $\,\theta^o\,$ is uniform.
This holds true when the initial state is static.
We follow  step by step the proof of Nowacki \cite{N:GTT} and put in evidence any difference when it will appear.

A modified version of energy balance is needed.  
It follows by substituting the virtual increments by the real increments
$$\delta u_\alpha=\frac{\partial u_\alpha}{\partial t}\,dt=v_\alpha\,dt  \,,  \qquad \delta u_{\alpha, \,M} = \dot u_{\alpha, \,M} \,dt \, , \quad \dots  $$
in the principle of virtual work
\begin{equation}    \label{eq:PVW}  
\int_{V^o} \Big( f^1_\alpha - \rho_o \ddot u_\alpha \Big) \delta u_\alpha \, dV + \int_{S^o}  \tilde K_\alpha \, \delta u_{\alpha} \, dS 
\,=\, \int_{V^o} \, K^1_{M\alpha} \, \delta u_{\alpha, \,M} \, dV   \, .
\end{equation}  

Thus the fundamental energy equation

\begin{equation}    \label{eq:EnEq}  
\int_{V^o} \Big( f^1_\alpha - \rho_o \dot v_\alpha \Big) v_\alpha \, dV + \int_{S^o} \tilde K_\alpha \, v_\alpha \, dS 
\,=\, \int_{V^o}  K^1_{M\alpha} \,\dot u_{\alpha, \,M}  \, dV  
\end{equation}

is obtained, where we substitute the constitutive relations (\ref{eq:K1KalpaG}).  Hence
\begin{eqnarray}    \label{eqnarray:EnEq2}  
\int_{V^o} \Big( f^1_\alpha - \rho_o \dot v_\alpha \Big) v_\alpha \, dV + \int_{S^o} \tilde K_\alpha \, v_\alpha \, dS 
\qquad \qquad \qquad \qquad \qquad \qquad\nonumber \\
\,=\, \int_{V^o}  \Big( G_{M \alpha L \gamma} u_{\gamma, \, L} +  R_{LM \alpha} \phi^1_{, \, L} - \rho_o \Lambda_{M \alpha}\theta^1 \Big) \, \dot u_{\alpha, \,M}\, dV   \, ,
\end{eqnarray}  
thus
 \begin{equation}    \label{eq:EnEq23}  
\frac{d}{dt}\Big( {\mathcal W}+{\mathcal K} \Big)\,=\,
\int_{V^o} f^1_\alpha \,  v_\alpha \, dV \, + \, \int_{S^o} \tilde K_\alpha \, v_\alpha \, dS \,+\,
 \int_{V^o}  \Big( \rho_o \Lambda_{M \alpha}\theta^1\,-\,R_{LM \alpha} \phi^1_{, \, L}  \Big) \, \dot u_{\alpha, \,M}\, dV   \, ,
\end{equation}  
where $\,{\mathcal W}\,$ is the work of deformation and $\,{\mathcal K}\,$ is the kinetic energy:
\begin{equation}      \label{eq:Wdef}  
	{\mathcal W}\,=\, \frac{1}{2}\, \int_{V^o} G_{M \alpha L \gamma} \, u_{\alpha, \,M} \, u_{\gamma, \, L} \, dV   \, ,
\qquad 	{\mathcal K}\,=\, \frac{1}{2}\, \int_{V^o} \rho_o \, v_\alpha v_\alpha \, dV   \, .
\end{equation}
Now, to eliminate the term 
$\; \int_{V^o} \rho_o \Lambda_{M \alpha}\theta^1\,  \dot u_{\alpha, \,M}\, dV   \, ,$ we multiply by $\, \theta^1\,$ the heat-conduction equation (\ref{eqnarray:N111}),  where $\,\dot \eta^o=0$,
 and integrate over $\,V^o$; after simple transformations we obtain
\begin{eqnarray}  \label{eqnarray:aabbcc}  
  \int_{V^o} \, \rho_o \, \theta^1\, \Lambda_{M \alpha} \, \dot u_{\alpha, \,M}\, dV  \,= 
  \frac{\kappa^E_{ML}}{\theta^o} \int_{S^o} \theta^1 \phi^1_{, \, L} N_M \, dS  \,+\, \qquad \qquad \qquad \nonumber \\ 
  \,+\,   \frac{\kappa_{L}}{\theta^o} \int_{S^o} \theta^1 N_L \, dS  
  \,+\,   \frac{\kappa_{ML}}{\theta^o} \int_{S^o} \theta^1 \theta^1_{, \, L} N_M \, dS  \,+\, \frac{\kappa_{ML\alpha}}{\theta^o} \int_{S^o} \theta^1 u_{\alpha, \,L} N_M \, dS  \qquad \\
  \,+\,   P_L \int_{V^o} \, \rho_o \,\theta^1 \dot \phi^1_{, \, L} \, dV   
  \,+\,   \frac{1}{\theta^o}  \int_{V^o} \, \rho_o \,\theta^1 \gamma^1 \, dV  
  \,-\,   \frac{d}{dt}{\mathcal P} \,- \,\big(\, \chi \,+\, \chi_\theta \,+\, \chi_\phi\,+\,\chi_u \big) \,,  \nonumber  \\\nonumber
\end{eqnarray}
where
\begin{equation}  \label{eq:Pi}
{\mathcal P} \,=\,  \frac{\alpha}{2\theta^o} \int_{V^o}\,\rho_o \theta^1\,\theta^1 \, dV \, ,
\end{equation}  
\begin{eqnarray}  \label{eqnarray:chi}  \nonumber
\chi_\phi = \frac{\kappa^E_{ML}}{\theta^o} \int_{V^o} \theta^1_{, \, M} \phi^1_{, \, L}  dV \,, \quad
\chi  = \frac{\kappa_{M}}{\theta^o} \int_{V^o} \theta^1_{, \, M} \theta^1  \, dV \,, \; \nonumber \\
\chi_\theta = \frac{\kappa_{ML}}{\theta^o} \int_{V^o} \theta^1_{, \, M} \theta^1_{, \, L}  dV \,,
\quad  \chi_u =  \frac{\kappa_{ML\alpha}}{\theta^o} \int_{V^o} \theta^1_{, \, M} u_{\alpha, \, L}  dV \,.
\end{eqnarray}
Note that this equation differs from the corresponding Eq.$(25)$ in \cite{N:GTT} by the terms $\,\chi_\phi$, $\;\chi\,$ and $\;\chi_u$.
Now, substituting (\ref{eqnarray:aabbcc})  into (\ref{eq:EnEq23}), we are lead to the equation
 \begin{eqnarray}    \label{eqnarray:subst50into48}  
\frac{d}{dt}\Big( {\mathcal W}+{\mathcal K}+{\mathcal P} \Big)
 \,+ \,\big(\, \chi \,+\, \chi_\theta \,+\, \chi_\phi\,+\,
 \chi_u\,\big) \,=\,
\int_{V^o} f^1_\alpha \,  v_\alpha \, dV \, + \, \int_{S^o} \tilde K_\alpha \, v_\alpha \, dS \,+\, \nonumber \\
+\, \frac{\kappa^E_{ML}}{\theta^o} \int_{S^o} \theta^1 \phi^1_{, \, L} N_M \, dS 
  \,+\,   \frac{\kappa_{L}}{\theta^o} \int_{S^o} \theta^1 N_L \, dS  
  \,+\,   \frac{\kappa_{ML}}{\theta^o} \int_{S^o} \theta^1 \theta^1_{, \, L} N_M \, dS  \,+ \qquad \\   
  \,+\,   \frac{1}{\theta^o}  \int_{V^o} \, \rho_o \,\theta^1 \gamma^1 \, dV  
  \,-\,   \int_{V^o} \,\Big(\, R_{LM \alpha} \phi^1_{, \, L} \, \dot u_{\alpha, \,M}\,-\,\rho_o  P_M  \,\theta^1 \dot \phi^1_{, \, M}\,\Big) \, dV   \, . \nonumber 
\end{eqnarray}  
  To eliminate the term 
  $$\,\int_{V^o} \Big(\, R_{LM \alpha} \phi^1_{, \, L} \, \dot u_{\alpha, \,M}\,-\,\rho_o  P_M  \,\theta^1 \dot \phi^1_{, \, M}\,\Big) dV \,$$  
  in Eq.(\ref{eqnarray:subst50into48}) we substitute the constitutive relations (\ref{eq:D1RLP}) into the time-derivative of the equation of the electric field (\ref{eq:D1}) with $\,\rho_E=0\,$.
  Multiplying  the obtained equation by $\,\phi^1\,$ and integrating over the region of the body,
  we obtain
  
\begin{equation}      \label{eq:Dp1=0}
	 \int_{S^o} \,\dot \Delta_{M} \phi^1 N_M \, dV \,+\,  
	 \int_{V^o} \,\dot \Delta_{M} W^1_M \, dV\,=\,0 \,.
\end{equation}

Using the relations (\ref{eq:D1RLP}) and (\ref{eq:Dp1=0}), after simple transformations we obtain
\begin{eqnarray}                 \label{eqnarray:dotDp1=0}
\int_{V^o} \dot \Delta_{L} W^1_L \,dV \,=\, \qquad \qquad \qquad   \qquad \qquad \qquad  \nonumber \\
=\, \int_{V^o} \Big( R_{LM \alpha}  \dot u_{\alpha, \,M} W^1_L \,+\,
 L_{LM}\dot W^1_{M}W^1_L \,+\, \rho_o P_L \frac{d}{dt}\big(\theta^1 W^1_L \big)
 \,-\, \rho_o P_L \theta^1 \dot W^1_L \Big) \,dV \,= \nonumber \\
=\, -  \int_{S^o} \dot \Delta^1_{L} N_L \phi^1 \, dS \,,\qquad \qquad \qquad   \qquad \qquad \qquad  \nonumber
\end{eqnarray}
from which
\begin{eqnarray}                 \label{eqnarray:fromwhich}
\int_{V^o} \,\Big( R_{KM \alpha}  \dot u_{\alpha, \,M} W^1_K
 \,-\, \rho_o P_K \theta^1 \dot W^1_K \Big) \,dV \,= \qquad \qquad \qquad \qquad \qquad \qquad \qquad  \nonumber \\
=\, - \, \int_{S^o} \, \dot \Delta^1_{K} N_K \phi^1 \, dS \,-\, \frac{d}{dt}\,{\mathcal E}
\,-\,\frac{d}{dt}\Big(\rho_o P_K \int_{V^o}\theta^1 W^1_K \,dV \Big)  \nonumber \\
\end{eqnarray}
where
\begin{equation}    \label{eq:MatE}  
{\mathcal E}	\,=\,\frac{1}{2}\,L_{KM} \int_{V^o} W^1_{M}W^1_K \,dV \,.
\end{equation}
In view of Eqs.(\ref{eqnarray:subst50into48}) and (\ref{eqnarray:fromwhich}), we arrive at the modified energy balance
 \begin{eqnarray}    \label{eqnarray:ModifEnBal}  
\frac{d}{dt}\Big( {\mathcal W}+{\mathcal K}+{\mathcal P+{\mathcal E}+
\rho_o P_K \int_{V^o}\theta^1 W^1_K \,dV \Big)}
 \,+ \,\big(\, \chi \,+\, \chi_\theta \,+\, \chi_\phi\,+\, \chi_U \,\big) \,=\, \nonumber \\
=\,\int_{V^o} f^1_\alpha \,  v_\alpha \, dV \, + \, \int_{S^o} \tilde K_\alpha \, v_\alpha \, dS \,+\,\qquad \qquad \qquad \qquad \nonumber \\
 \,+\,   \frac{\kappa^E_{ML}}{\theta^o} \int_{S^o} \theta^1 \phi^1_{, \, L} N_M \, dS 
  \,+\,   \frac{\kappa_{L}}{\theta^o} \int_{S^o} \theta^1 N_L \, dS  
  \,+\,   \frac{\kappa_{ML}}{\theta^o} \int_{S^o} \theta^1 \theta^1_{, \, L} N_M \, dS  \,+ \qquad \\   
  \,+\,   \frac{1}{\theta^o}  \int_{V^o} \, \rho_o \,\theta^1 \gamma^1 \, dV  
  \, - \, \int_{S^o} \, \dot \Delta^1_{K} N_K \phi^1 \, dS  \, . \nonumber 
\end{eqnarray}  
The energy balance (\ref{eqnarray:ModifEnBal}) makes possible the proof of the uniqueness of the solution.

We assume that two distinct solutions $\,(u'_i,\; \phi^1{'},\; \theta^1{'})\,$
and $\,(u_i'',\; \phi^1{''},\; \theta^1{''})\,$ satisfy Eqs.(\ref{eq:k1})-(\ref{eq:N1})
and the appropriate boundary and initial conditions.  Their difference
$$\,(\hat u_i=u'_i-u_i'',\quad \hat \phi=\phi^1{'}-\phi^1{''},\quad \hat \theta=\theta^1{'}=\theta^1{''})\,$$
 therefore satisfies the homogeneous equations 
 (\ref{eq:k1})-(\ref{eq:N1}) and the homogeneous boundary and initial conditions.
 Equation (\ref{eqnarray:ModifEnBal}) holds for 
 $\,(\hat u_i,\; \hat \phi,\; \hat \theta)$.
 
 In view of the homogeneity of the equations and the boundary conditions, the right-hand side of Eq.(\ref{eqnarray:ModifEnBal})  vanishes.
 Hence
 \begin{eqnarray}    \label{eqnarray:ModifEnBalDIFFERENCE}  
\frac{d}{dt}\Big( {\mathcal W}+{\mathcal K}+{\mathcal P+{\mathcal E}+
\rho_o P_K \int_{V^o}\theta^1 W^1_K \,dV \Big)}
 \,= \,-\,\big(\, \chi \,+\, \chi_\theta \,+\, \chi_\phi\,+\,\chi_u\,\big) \,\leq \, 0 \,, \nonumber \\
\end{eqnarray}  
where the last inequality is true since by (\ref{eq:Qflow}),
(\ref{eqnarray:chi}) and (\ref{eq:RQ1}) we have
\begin{equation}  \label{eq:eq33}  
-\big(\,\chi \,+\, \chi_\theta \,+\, \chi_\phi\,+\, \chi_u \, \big)
\,=\,\frac{1}{\theta^o} \int_{V^o} Q^1_M\Theta^1_M \,dV  \, .
\end{equation}
The integral in the left-hand side of Eq.(\ref{eqnarray:ModifEnBalDIFFERENCE}) vanishes at the initial instant, since the functions 
$\,\hat u_i,\, \hat \phi,\, \hat \theta\,$ satisfy the homogeneous initial conditions.
On the other hand, by the inequality in (\ref{eqnarray:ModifEnBalDIFFERENCE}) the left-hand side is either negative or zero.

Now we assume  $(i-iii)$ below; note that $\,(iii)\,$ is the sufficient condition of J. Ignaczak, written in \cite{N:GTT} on pages 176-177. 

$(i)$ The initial deformation $\,y^o_\alpha\,$ realizes that the tensor $\, G_{M \alpha L \gamma}\,$ is positive-definite, so that $\,{\mathcal W}\geq 0\,$ by (\ref{eq:Wdef}).

$(ii)$  The tensor $\, L_{KN} \,$ is positive-definite so that,  by (\ref{eq:MatE}), $\,{\mathcal E}\geq 0$.

$(iii)$  $\, L_{IJ}\,$ is a known positive-definite symmetric tensor, $\,g_I=\rho_oP_I\,$ is a vector, and $\,c=\rho_o\alpha/2\theta^o>0$;  consider the function 
$$A(\theta^1, \, W_L)=\ (\theta^1)^2+2\theta^1 g_IW^1_I+L_{IJ}W^1_IW^1_J$$
$A$ is nonnegative for every real pair $\,(\theta^1,\,W^1_k)\,$, provided
$$|g_I|\leq c \lambda_m$$
where $\, \lambda_m\,$ is the smallest positive eigenvalue of the tensor $\,L_{IJ}$.


Under these three assumptions, (\ref{eqnarray:ModifEnBalDIFFERENCE}) implies
$$\hat{u}_{i, \,L}=0,\qquad \hat{\theta}=0,\qquad  \hat W_{L}=0\,,$$

which imply the uniqueness of the solutions of the incremental thermoelectroelastic equations, i.e.,
$$u'_i=u_i'',\qquad \theta^1{'}=\theta^1{''},\qquad  W^1_I{'}=W^1_I{''}\,.$$
Moreover, from the constitutive relations we have that
$$K^1_{I\alpha}{'}=K^1_{I\alpha}{''},\qquad \Delta^1_L{'}=\Delta^1_L{''},\qquad  \eta^1{'}=\eta^1{''} \, .$$

\section{On the generalized Hamilton's principle}    \label{section:genhampri} 
We define the free energy, electric enthalpy, and potential of the heat flow respectively by
\begin{equation}  \label{eq:psi1}
	\psi^1\,=\, \frac{1}{2} G_{M \alpha L \gamma} u_{\alpha, \, M}  u_{\gamma, \, L} +  R_{LM \alpha} \phi^1_{, \, L} u_{\alpha, \, M}
	 - \rho_o \theta^1 \Big[\Lambda_{M \alpha} u_{\alpha, \, M} - P_{M}\phi^1_{, \, M}+\frac{\alpha}{2}\theta^1\Big] \, , 
\end{equation}

\begin{equation}
	H^1\,=\, \psi^1 - \frac{1}{2}L_{AB}W^1_AW^1_B  \,=\, \psi^1 
	- \frac{1}{2}L_{AB}\Phi^1_{, \,A} \Phi^1_{, \,B} \, , 
\end{equation}
and
  \begin{equation}    \label{eq:Qpotent} 
\Gamma =-\Big(\,\kappa_{MN\alpha}u_{\alpha, \,N}\theta^1_{,\,M}
               + \frac{1}{2}\kappa_{MN} \theta^1_{,\,M}\theta^1_{,\,N}  
               + \kappa^E_{MN} \theta^1_{,\,M}\phi^1_{,\,N} 
               + \kappa_{M} \theta^1 \theta^1_{,\,M}\,\Big)  \, .
 \end{equation}

Whence
\begin{equation}
	     \frac{\partial H^1}{\partial u_{\alpha, \, M}}\,=\,K^1_{M \alpha} \, ,  \quad 
	   	\frac{\partial H^1}{\partial W^1_L}\,=\, - \Delta^{1}_L \, ,         \quad  
	   		\frac{\partial H^1}{\partial \theta}\,=\, - \rho_o \eta^{1} \, ,
\end{equation}
  \begin{equation}    \label{eq:gfrac} 
Q^1_{M} \,=\, \frac{\partial \Gamma}{\partial \theta^1_{,M}} \, .
 \end{equation}
Lastly we define two functionals
\begin{equation}
	\Pi\,=\, \int_{V^o} \Big( H^1 + \rho_o\eta^1 \theta^1- f^1_\alpha u_\alpha \Big) dV \,-\,
	             \int_{S^o} \Big( \tilde K^1_\alpha u_\alpha - \tilde \Delta^1 \phi^1 \Big) dS  
\end{equation}
and
 \begin{equation}  \label{eq:psi2fu}
	\Psi
\,=\,\int_{V^o} \Big( \Gamma - \rho_o (\eta^1 \theta^o \dot \theta^1
 +\eta^1 \dot \theta^o \theta^1 	+ \eta^o \theta^1 \dot \theta^1 + \gamma^1\theta^1) \Big) dV 
 \,+\, \int_{S^o} \theta^1 \tilde Q \, dS  \, ,  
\end{equation}
Eqs.(\ref{eq:psi1})-(\ref{eq:psi2fu}) generalize Eqs.\cite[(36)-(38)]{N:GTT}.

\bigskip

The generalized Hamilton's principle
has the form
\begin{equation}               \label{eq:deltaIeII}
\delta \, \int^{t_2}_{t_1} \Big( {\mathcal K} - \Pi \Big) \, dt \,=\, 0  \, , \qquad
\delta \, \int^{t_2}_{t_1} \Psi \, dt \,=\, 0  \,                \end{equation}
The virtual  processes 
$$(\delta u_\alpha,\, \delta \theta^1,\, \delta \phi^1)$$
of the body must be compatible with the conditions restricting the process of the body.
Moreover the virtual processes  must satisfy the conditions
$$\delta u_\alpha({\bf x},\,t_1)=\delta u_\alpha({\bf x},\,t_2)=0, \; \delta \theta^1({\bf x},\,t_1)=\delta \theta^1({\bf x},\,t_2)=0, \; \delta \phi^1({\bf x},\,t_1)=\delta \phi^1({\bf x},\,t_2)=0.$$

Hence, performing the variations in the second of Eqs.(\ref{eq:deltaIeII}) and observing that
\begin{equation}
	\delta H^1\,=\,K^1_{M \alpha} \delta u_{\alpha, \, M}- \rho_o \eta^{1} \delta \theta^1 + \Delta^{1}_L  \delta	\Phi^1_{, \,L}  \, ,
\end{equation}
and
\begin{eqnarray}  \nonumber
\int^{t_2}_{t_1} \Big( {\mathcal K} - \Pi \Big) \, dt \,=\,
\qquad  \qquad  \qquad  \qquad  \qquad  \qquad  \qquad \qquad  \qquad  \qquad  \qquad  \qquad   \nonumber \\
\,=\,  \int^{t_2}_{t_1} dt \,\Big[
 \int_{V^o} \Big( \frac{\rho_o}{2} \dot u_\alpha \dot u_\alpha \,-\, H^1 \,-\, \rho_o \eta^1 \theta^1 \,+\, f^1_\alpha u_\alpha \Big) dV
 \,+\,\int_{S^o} \Big( \tilde K^1_\alpha u_\alpha - \tilde \Delta^1 \phi^1 \Big) dS   \, \Big],  \quad 
\end{eqnarray}
we have
\begin{eqnarray}  \nonumber
\delta \, \int^{t_2}_{t_1} \Big( {\mathcal K} - \Pi \Big) \, dt \,=\,  
\int^{t_2}_{t_1} dt \,\Big[ \int_{V^o} \Big( - \rho_o \ddot u_\alpha \delta u_\alpha \,-\, K^1_{M \alpha} \delta u_{\alpha, \, M} - \Delta^{1}_L  \delta \Phi^{1}_{,\,L} \,+\, f^1_\alpha \delta u_\alpha \Big) dV 
 \qquad  \nonumber \\ 
 \qquad \qquad \,+\,\int_{S^o} \Big( \tilde K^1_\alpha \delta u_\alpha - \tilde \Delta^1 \delta \phi^1 \Big) dS \,\Big] \, .   \qquad \qquad \qquad \qquad 
\end{eqnarray}
Hence by the identities    
\begin{eqnarray}  \nonumber
-K^1_{L \alpha} \big(  \delta u_{\alpha}\big)_{,\,L} \,=\, -\big( K^1_{L \alpha} \delta u_{\alpha} \big)_{,\,L}  \,+\,
\big( K^1_{L \alpha \,,\,L}\big) \delta u_{\alpha}\, , \nonumber \\
\Delta^1_{L} \big(  \delta \phi^1\big)_{,\,L} \,=\, \big( \Delta^1_{L} \delta \phi^1 \big) _{,\,L} \,-\,
\big( \Delta^1_{L,\,L}\big) \delta \phi^1\, ,
\end{eqnarray}
we have 
\begin{eqnarray}  \nonumber
\delta \, \int^{t_2}_{t_1} \Big( {\mathcal K} - \Pi \Big) \, dt \,=\, 
\int^{t_2}_{t_1}  dt \,\Big[  \int_{V^o} \Big[ \Big( - \rho_o \ddot u_\alpha \delta u_\alpha \,+\, K^1_{M \alpha, \, M}+f^1_\alpha \Big)  \delta u_{\alpha} 
\,+\,  \Delta^1_{M, \, M} \delta \phi^1  \Big] dV  \nonumber \\
\qquad \,+\, \int_{S^o}\Big( - K^1_{M \alpha} \delta u_\alpha N_M dS -\Delta^1_{M} \delta \phi^1 N_M \Big) dS
 \,+\, \int_{S^o} \Big( \tilde K^1_\alpha \delta u_\alpha - \tilde \Delta^1 \delta \phi^1 \Big) dS\Big]\, . \qquad \qquad
\end{eqnarray}

Thus we have
\begin{eqnarray}  \label{eqnarray:FIfin} \nonumber
\int^{t_2}_{t_1}  dt \, \Big[ \, \int_{V^o} \Big( - \rho_o \ddot u_\alpha \,+\, K^1_{M \alpha, \, M}+f^1_\alpha \Big)  \delta u_{\alpha}  \,dV 
 \,+\, \int_{V^o}    \Delta^1_{M, \, M} \delta \phi^1  \,dV \nonumber \\
\,+\, \int_{S^o}\Big( \tilde K^1_\alpha  - K^1_{M \alpha}  N_M \Big)\delta u_\alpha dS 
 \,-\, \int_{S^o}  \Big( \tilde \Delta^1 + \Delta^1_{M} N_M \Big) \delta \phi^1 \, dS \, \Big] \,=\, 0  \, .
\end{eqnarray}

Since the variations $\delta u_\alpha$ and $\delta \phi^1$ are arbitrary, Eq.(\ref{eqnarray:FIfin}) is equivalent to the equations governing the incremental motion and electric field, completed by the appropriate boundary conditions.  These equations and boundary conditions coincide with those written above.

------------------------------------------------------------------------------------------

Next we perform the required variation in the second of Eqs.(\ref{eq:deltaIeII}) 
by observing that 
\begin{eqnarray}      \label{eqnarray:deltatau}
	\delta \Gamma\,=\,\frac{\partial \Gamma}{\partial u_{\alpha, \, N}} \delta u_{\alpha, \, N}
	\,+ \,\frac{\partial \Gamma}{\partial \theta^1_{,\,L}} \delta \theta^1_{,\,L}
	\,+ \,\frac{\partial \Gamma}{\partial \phi^1_{,\,L}} \delta \phi^1_{,\,L}
	\,+\,\frac{\partial \Gamma}{\partial \theta^1} \delta \theta^1  \nonumber  \\
\,=\,-\kappa_{MN\alpha}\theta^1_{,\,M}\delta u_{\alpha, \, N}
	\,+ \,Q^1_L \delta \theta^1_{,\,L}
	\,- \,\kappa^E_{MN} \theta^1_{,\,M} \delta \phi^1_{,\,L}
	\,-\, \kappa_{M} \theta^1_{,\,M} \delta \theta^1     \, .
\end{eqnarray}

By (\ref{eq:psi2fu}) we have
\begin{eqnarray}               \label{eqnarray:deltaII}
\delta \,\int^{t_2}_{t_1} \Psi \,dt \,=\, \nonumber \qquad \qquad \qquad\qquad \qquad \qquad \qquad \qquad  \\
=\, \int^{t_2}_{t_1}dt \Big[ \int_{V^o} \Big( \delta \Gamma 
- \rho_o \eta^1( \theta^o  \delta \dot \theta^1 +\dot \theta^o \delta \theta^1)
- \rho_o \eta_o(\theta^1 \delta\dot \theta^1+\dot \theta^1 \delta \theta^1) 
- \rho_o \gamma^1 \delta \theta^1)  \Big) dV\nonumber \\
 \,+\,
	             \int_{S^o} \delta \theta^1 \tilde Q\, dS \Big] \nonumber  \\
=\, \int^{t_2}_{t_1}dt \Big[ \int_{V^o} \Big(  -\kappa_{ML\alpha}\theta^1_{,\,M}\delta u_{\alpha, \, L}
	\,+ \,Q^1_L \delta \theta^1_{,\,L} 
	\,- \,\kappa^E_{ML} \theta^1_{,\,M} \delta \phi^1_{,\,L}
	\,-\, \kappa_{M} \theta^1_{,\,M} \delta \theta^1 \nonumber \\
+ \rho_o[\dot \eta^1 \theta^o \delta \theta^1 - \dot{\overline{(\eta^1 \theta^o \delta \theta^1)}}]
+ \rho_o[\dot \eta^o \theta^1 \delta \theta^1 - \dot{\overline{(\eta^o \theta^1 \delta \theta^1)}}]
- \rho_o \gamma^1 \delta \theta^1  \Big) dV  \nonumber \\
	\,+\, \int_{S^o} \delta \theta^1 \tilde Q\, dS \Big] \, . \qquad \qquad \qquad \qquad 
 \end{eqnarray}
Note that
\begin{equation}               \label{eq:ideta}
\int^{t_2}_{t_1}  \dot{\overline{(\eta^\nu \theta^\tau \delta \theta^1 )}} dt \,=\, \Big[ \eta^\nu \theta^\tau \delta \theta^1\Big]^{t_2}_{t_1}\,=\,0   \,, \qquad \nu, \tau=0,\,1,
    \end{equation}
 since  $\, \delta \theta^1=0\,$  at $\,t_1\,$ and $\,t_2$.  Also by using the identity 
 \begin{equation}               \label{eq:ident}
(a_L\,b)_{,\,L} \,=\, a_{L,\,L}\,b\,+\,a_L\,b_{,\,L}     
    \end{equation}
we obtain
\begin{eqnarray}               \label{eqnarray:deltaIIbb}
\delta \, \int^{t_2}_{t_1} \Psi \, dt\,=\,\qquad \qquad \qquad \qquad\qquad \qquad \qquad \qquad\nonumber  \\
=\, \int^{t_2}_{t_1}dt \Big[ \int_{V^o} \Big( \,Q^1_{L,\,L}  
	 + \rho_o[ \dot \eta^1 \theta^o+\dot \eta^o \theta^1 - \gamma^1 ]  \Big)\delta \theta^1  dV 
	\,-\,  \int_{S^o}  \Big(Q^1_L N_L -\tilde Q \Big)  \delta \theta^1 \, dS \Big]  \nonumber \\
-\, \int^{t_2}_{t_1}dt \Big[ \int_{V^o} \Big(  \kappa_{ML\alpha}\theta^1_{, \, M} \delta u_{\alpha, \,L}
	+\kappa^E_{IJ} \theta^1_{,\,I} \delta \phi^1_{,\,J}
	+ \kappa_{L} \theta^1_{,\,L} \delta \theta^1 \Big)\,dV \Big] 
    \end{eqnarray}
with
\begin{eqnarray}               \label{eqnarray:deltaIIbbcc}
\int_{V^o} \kappa_{ML\alpha}\theta^1_{, \, M} \delta u_{\alpha, \, L}\,dV
\,=\,\kappa_{ML\alpha}\Big[- \int_{V^o} \theta^1_{, \, ML} \delta u_{\alpha}\,dV \,+\,\int_{S^o} \theta^1_{, \, M} N_L \delta u_{\alpha}\,dS \Big]\, ,
    \end{eqnarray}
\begin{eqnarray}               \label{eqnarray:deltaIIbbccvv}
\int_{V^o} \kappa^E_{ML} \theta^1_{,\,M} \delta \phi^1_{,\,L}\,dV
\,=\,\kappa^E_{ML} \Big[- \int_{V^o} \theta^1_{, \,ML} \delta \phi^1\,dV \,+\, \int_{S^o} \theta^1_{, \, M}  N_L \delta \phi^1\,dS \Big]\, .
    \end{eqnarray}
Hence, by performing the variation (\ref{eqnarray:deltaIIbb}) with
the variations $\, \delta u_{\alpha}, \delta \phi^1\,$ that vanish, and with 
 $\,\delta \theta^1\,$ arbitrary, we obtain that  (\ref{eqnarray:deltaIIbb}) reduces to
\begin{eqnarray}               \label{eqnarray:deltaIIbbred}
\delta \, \int^{t_2}_{t_1} \Psi \, dt \,=\,
 \int^{t_2}_{t_1}dt \Big[ \int_{V^o} \Big( \,Q^1_{L,\,L}  - \kappa_{L} \theta^1_{,\,L} 
	 + \rho_o[ \dot \eta^1 \theta^o+\dot \eta^o \theta^1 - \gamma^1 ]  \Big)\delta \theta^1  \, dV  \nonumber  \\
\,-\,\int_{S^o} \Big(Q^1_L N_L -\tilde Q \Big) \delta \theta^1 \, dS \Big]   \, .   \end{eqnarray}

Thus $\,(i)\,$ {\it the variational equation (\ref{eq:deltaIeII})$_2$
 performed with 
\begin{equation}
	            \delta u_\alpha \,=\,0\,=\,\delta \phi^1
\end{equation}
is equivalent to the entropy balance
\begin{equation} \label{eq:entrbal}
Q^1_{L,\,L}+ \rho_o(\dot \eta^1 \theta^o + \dot{\eta}^o \theta^1 -\gamma^1 )\,=\,0 \, \end{equation}
and the boundary condition for the heat flow 
\begin{equation} \label{eq:bchf}
Q^1_L N_L \,=\, \tilde Q \,, \qquad ( \, {\bf x} \in S \, )
\end{equation}
 if and only if 
 
\begin{equation}
	\kappa_L\,=\,0  \, .
\end{equation}
}
 
 Alternatively, 
by performing the variation (\ref{eq:deltaIeII})$_2$ with
all the variations $\, \delta u_{\alpha}, \; \delta \phi^1, \; \delta \theta^1\,$ arbitrary, we deduce that

 $\,(ii)\,$  {\it the variational equation (\ref{eq:deltaIeII})$_2$
is equivalent to the entropy balance (\ref{eq:entrbal})
and the boundary condition for the heat flow (\ref{eq:bchf})
 if and only if} $$\kappa_L=0, \quad \kappa^E_{ML}=0, \quad \kappa_{ML\alpha}=0\,.$$
 
\section{Theorem of Reciprocity of Work}   \label{section:RecWork} 
Next we extend the theorem of reciprocity of work following some steps in \cite{N:GTT} on pages 179-182, where it is referred to linear thermoelectroelasticity in a natural configuration.  Here there are some essential changes imposed by the presence of the initial fields.
We assume that the body is homogeneous and moreover  that the initial state is static, so that in particular $\,\dot \theta^o=0$, $\,\dot \eta^o=0$.
Here we do not assume that  $\, \theta^o\,$ is uniform.

The Laplace transform
of functions $\,\nu=\nu({\bf x},\, t)\,$,
\begin{equation}     \label{eq:LaplTranDEF}
	\overline \nu({\bf x},\, p)\,=\, \int_0^{\infty}\, e^{-pt}\nu({\bf x},\, t)\,dt \, ,
\end{equation}
	 will be used below.

Consider two sets of causes  $\,{\mathcal A}^1$,  $\,{\mathcal A}^1{'}\,$ for incremental processes, and respective effects  $\,(u_\alpha,\, \phi,\, \theta)$,  $\,(u'_\alpha,\, \phi',\, \theta')$.
Starting 
from the equations of motion
\begin{eqnarray}                    \label{eqnarray:KLkkk}
	K^1_{L\alpha,L}+\rho_o f_\alpha=\rho_o\ddot u_\alpha \, ,  \\
	K^1{'}_{\!\!\!\!\!L\alpha,L}+\rho_o f'_\alpha=\rho_o\ddot u'_\alpha \, ,
	\end{eqnarray}
	taking their Laplace transform, multiplying each by $\,\overline{\theta^o}$,
	then multiplying the first by $\,\overline u{'}_{\alpha}\,$ and the second by
	$\,\overline u{}_{\alpha}$,   and making the difference of their integrals over the instantaneous region $V$, assuming that the initial conditions for the displacements are homogeneous,
we obtain the integral equation
\begin{eqnarray}         \label{eqnarray:intKLkkk}
\int_{V^o} \overline{\theta^o} \Big( \overline F_\alpha \overline u{'}_{\alpha}-\overline F{'}_\alpha \overline u_{\alpha} \Big)\,dV \,+\,
\int_{V^o} \overline{\theta^o} 
\Big(\overline{K^1_{\!\!L \alpha,\,L}} \overline{u'_{\alpha}}-\overline{K^1{'}_{\!\!\!\!\!\!L \alpha,\,L}} \overline{u_{\alpha}} \Big)  \,dV
\,=\,0\,, 
\end{eqnarray}
where $\,F_\alpha= \rho_o f_\alpha \, , \; F{'}_\alpha= \rho_o f'_\alpha$.
Now, by the identity (\ref{eq:ident}) and the divergence theorem, we have
   \begin{eqnarray}        \nonumber
\int_{V^o} \overline{\theta^o} 
\Big(\overline{K^1_{\!\!L \alpha,\,L}} \overline{u'_{\alpha}}-\overline{K^1{'}_{\!\!\!\!\!\!L \alpha,\,L}} \overline{u_{\alpha}} \Big)  \,dV
\,=\,
\int_{S^o} \overline{\theta^o} 
\Big(\overline{K^1_{\!\!L \alpha}} \overline{u'_{\alpha}}-\overline{K^1{'}_{\!\!\!\!\!\!L \alpha}} \overline{u_{\alpha}} \Big)N_L  \,dS
\nonumber \\
\,-\,
\int_{V^o} \Big(\overline{K^1_{\!\!L \alpha}}(\overline{\theta^o} \overline{u'_{\alpha}})_{,\,L}
-\overline{K^1{'}_{\!\!L \alpha}}(\overline{\theta^o} \overline{u_{\alpha}})_{,\,L} \Big)  \,dV \,, \qquad  \nonumber 
   \end{eqnarray}         
 hence
   \begin{eqnarray}         \label{eqnarray:intKLkkkiddt}
\int_{V^o} \overline{\theta^o} \Big(\overline{K^1_{\!\!L \alpha,\,L}} \overline{u'_{\alpha}}-\overline{K^1{'}_{\!\!\!\!\!\!L \alpha,\,L}} \overline{u_{\alpha}} \Big)  \,dV
\,=\,
\int_{S^o} \overline{\theta^o} 
\Big(\overline{K^1_{\!\!L \alpha}} \overline{u'_{\alpha}}-\overline{K^1{'}_{\!\!\!\!\!\!L \alpha}}
 \overline{u_{\alpha}} \Big)N_L  \,dS \quad
\nonumber \\
\,-\,
\int_{V^o} (\overline{\theta^o})_{,\,L} \Big(\overline{K^1_{\!\!L \alpha}}\overline{u{'}_{\alpha}}
-\overline{K^1{'}_{\!\!L \alpha}} \overline{u_{\alpha}} \Big)  \,dV\,-\,
\int_{V^o} \overline{\theta^o} \Big(\overline{K^1_{\!\!L \alpha}}(\overline{u{'}_{\alpha}})_{,\,L}
-\overline{K^1{'}_{\!\!\!\!\!\!L \alpha}}(\overline{u_{\alpha}})_{,\,L} \Big)  \,dV  \,. 
 \end{eqnarray}
Hence by the latter equation and the constitutive relations (\ref{eq:K1KalpaG}), Eq.(\ref{eqnarray:intKLkkk})  becomes
\begin{eqnarray}         \label{eqnarray:intKLkkky}
\int_{V^o} \overline{\theta^o} \Big( \overline F_\alpha \overline u{'}_{\alpha}-\overline F{'}_\alpha \overline u_{\alpha} \Big)\,dV \,+\,
\int_{S^o} \overline{\theta^o} \Big( \overline K^1_{L\alpha} \overline u{'}_{\alpha}-\overline K^{1'}_{L\alpha} \overline u_{\alpha} \Big)N_L\,dS \nonumber \\
\,+\, \int_{V^o} \overline{\theta^o} \Big[\rho_o \Lambda_{L \alpha} \Big(\overline \theta^{1'} \overline u_{\alpha, \,L} - \overline \theta^1 \overline u'_{\alpha, \,L}\Big)
\,+\, R_{L N \gamma} \Big(
\overline{u_{\gamma, \, N}}\, \overline{{W^1}_{\!\!\!\!L}{'}}
\,-\,\overline{u{'}_{\!\!\!\gamma, \, N}}
\,\overline{W^1_{L}}\Big)\Big]  \,dV \nonumber\\
\,-\,
\int_{V^o} (\overline{\theta^o})_{,\,L} \Big(\overline{K^1_{\!\!L \alpha}}\overline{u'}_{\alpha}
-\overline{K^1_{\!\!L \alpha}{\!\!'}}\; \overline{u}_{\alpha} \Big)  \,dV
\,=\,0\,, 
\end{eqnarray}
 which is the analogue of Eq.$(54)$ in \cite{N:GTT}.  

-----------------------------------------------------------------------------------------

Next we shall make use of the heat-conduction equation (\ref{eq:entrbal}) for both the systems of loadings, rewritten in the form
\begin{equation}    \label{eq:N11}  
-\overline{\Big(\frac{1}{\theta^o}Q^1_{M, \, M}\Big)} -
 \rho_o \overline{\dot \eta^1}\,=\, -\rho_o\overline{\Big(\frac{\gamma^1}{\theta^o}\Big)} \, ,
\end{equation}  
since we have 
\begin{equation}
	\overline{\dot \eta^o} \,=\,0 \, .
\end{equation}
Hence by Eqs.(\ref{eq:Qflow}) and (\ref{eq:Eta1P}) we obtain
\begin{eqnarray}	   \nonumber
\Big(\kappa_{LN\alpha}\overline{\frac{u_{\alpha, \,NL}}{\theta^o}}
+\kappa^E_{MN}\overline{\frac{\phi^1_{, \,NM}}{\theta^o}} 
+ \kappa_{L} \overline{\frac{\theta^1_{, \, L}}{\theta^o}} 
+ \kappa_{MN} \overline{\frac{\theta^1_{,\,NM}}{\theta^o}}  	\Big)
 \qquad \qquad \qquad\qquad \qquad \qquad \nonumber \\
 \,-\,
p \rho_o \Big(\Lambda_{M \gamma} \overline u_{\gamma, \, M} -P_{M} 
\overline \phi^1_{, \, M} + \alpha \overline \theta^1)
\,=\,-\rho_o\overline{\Big(\frac{\gamma^1}{\theta^o}\Big)} \,.
\qquad \qquad \qquad
\end{eqnarray} 
Multiplying the latter by $\,\overline {\theta^o}\,$ we have
\begin{eqnarray}	
\overline {\theta^o}\Big(\kappa_{LN\alpha}\overline{\frac{u_{\alpha, \,NL}}{\theta^o}}
+\kappa^E_{MN}\overline{\frac{\phi^1_{, \,NM}}{\theta^o}} 
+ \kappa_{L} \overline{\frac{\theta^1_{, \, L}}{\theta^o}} 
+ \kappa_{MN} \overline{\frac{\theta^1_{, \, NM}}{\theta^o}}   	\Big) \nonumber \\
\,-\,
p \rho_o \overline {\theta^o} \Big(\Lambda_{M \gamma} \overline u_{\gamma, \, M} -P_{M} 
\overline \phi^1_{, \,NM} + \alpha \overline \theta^1)
\,=\,- \overline {\theta^o}\rho_o\overline{\Big(\frac{\gamma^1}{\theta^o}\Big)} \,.
\end{eqnarray}
Write the latter equality for both the states, multiply the first equation by 
$\,\overline {\theta^1}'\,$ and the second by $\,\overline {\theta^1}\,$ ; we obtain
\begin{eqnarray}	
\overline {\theta^1}' \overline {\theta^o}\Big(\kappa_{LN\alpha}\overline{\frac{u_{\alpha, \,NL}}{\theta^o}}
+\kappa^E_{MN}\overline{\frac{\phi^1_{, \,NM}}{\theta^o}} 
+ \kappa_{L} \overline{\frac{\theta^1_{, \, L}}{\theta^o}} 
+ \kappa_{MN} \overline{\frac{\theta^1_{, \, NM}}{\theta^o}}   	\Big)
 \nonumber \\
\,-\,
p \rho_o \overline {\theta^1}'\overline {\theta^o} \Big(\Lambda_{M \gamma} \overline u_{\gamma, \, M} -P_{M} 
\overline \phi^1_{, \,NM} + \alpha \overline \theta^1) 
\,=\,
- \overline{\theta^1{'}} \, \overline {\theta^o}\rho_o\overline{\Big(\frac{\gamma^1}{\theta^o}\Big)} \, ,
\end{eqnarray}
and 
\begin{eqnarray}	
\overline{\theta^1} \overline{\theta^o}
\Big(\kappa_{LN\alpha}\overline{\frac{u_{\alpha, \,NL}{'}}{\theta^o}}
+\kappa^E_{MN}\overline{\frac{\phi^1_{,\,MN}{'}}{\theta^o}} 
+ \kappa_{L} \overline{\frac{\theta^1_{, \, L}{'}}{\theta^o}} 
+ \kappa_{MN} \overline{\frac{\theta^1_{, \, MN}{'}}{\theta^o}}\Big)   \nonumber \\
\,-\,
p \rho_o \overline {\theta^1}\overline {\theta^o} \Big(\Lambda_{M \gamma} \overline u{'}_{\gamma, \, M} -P_{M} 
\overline \phi^{1}_{, \, M}{\!\!'} + \alpha \overline \theta^{1'}) 
\,=\,- \overline{\theta^1}\, \overline{\theta^o} \rho_o\overline{\Big(\frac{\gamma^{1}{'}}{\theta^o}\Big)} \,.
\end{eqnarray}

By taking the integral over $V$ of the difference between the last two equations, we obtain the analogue of Eq.$(57)$ in \cite{N:GTT}, that is,
\begin{eqnarray}	\label{eqnarray:577}	
\kappa_{LN\alpha} \int_{V^o} \overline {\theta^o}\Big(\overline {\theta^1}'\overline{\frac{u_{\alpha, \,NL}}{\theta^o}} -
\overline {\theta^1}\overline{\frac{u'_{\alpha, \,NL}}{\theta^o}}\Big)\,dV \,+\,
\kappa^E_{MN} \int_{V^o} \overline {\theta^o}
\Big(\overline {\theta^1{'}}\overline{\frac{\phi^1_{, \, MN}}{\theta^o}} -
\overline {\theta^1}\overline{\frac{\phi^1{'}_{\!, M}}{\theta^o}}\Big)\,dV \,+\, \nonumber \\
\kappa_{L} \int_{V^o} \overline {\theta^o}
\Big(\overline {\theta^1{'}}\overline{\frac{\theta^1_{, \, L}}{\theta^o}} -
\overline {\theta^1}\overline{\frac{\theta^1_{,\, L}{'}}{\theta^o}}\Big)\,dV \,+\,
\kappa_{MN} \int_{V^o} \overline {\theta^o}
\Big(\overline {\theta^1{'}}\overline{\frac{\theta^1_{, \, NM}}{\theta^o}} -
\overline {\theta^1}\overline{\frac{\theta^1{'}_{\!, M}}{\theta^o}}\Big)\,dV \,+\, \nonumber \\
+\, p \int_{V^o} \overline \rho_o{\overline{\theta^o}}
\Big[\overline {\theta^1{'}}\Big(-\Lambda_{M \gamma} \overline u{}_{\gamma, \, M} -P_{M} \overline{W^1{}_{\!\!\!\!M}}\Big) 
\,+\,
\overline {\theta^1} \Big(\Lambda_{M \gamma} \overline u{'}_{\gamma, \, M} +P_{M} \overline{W^1{'}_{\!\!\!\!M}}\Big) 
\Big]\,dV \,+\, \nonumber \\
\,+\, \int_{V^o} \rho_o {\overline{\theta^o}}  
\Big(\overline {\theta^1}\overline{\frac{\gamma^1{'}}{\theta^o}}
\,-\,\overline {\theta^1{'}} \overline{\frac{\gamma^1}{\theta^o}}
\Big)\,dV \,=\,0 \, . \nonumber \\
\end{eqnarray}

Finally, we make use of the equation for the electric field
\begin{equation}                  
	\overline {\Delta^1}_{L,L}\,=\,0 \, ,	\qquad \overline{\Delta^1}{'}_{L,L}\,=\,0 \, .
\end{equation}

Multiplying both by $\,\overline {\theta^o}$,  the first by $\,\overline \phi{'}$, the second by $\,\overline \phi$, subtracting the results and integrating over the region of the body, we obtain 
\begin{equation}                  
\int_{V^o}\Big(\overline{{\Delta^1}_{L,L}}\,(\overline {\theta^o}\, \overline {{\phi^1}{'}})
\,-\,\overline{{\Delta^1}{'}_{\!\!L,L}}(\overline {\theta^o}\, \overline {\phi^1}) \Big)dV \,=\,0 \, .
\end{equation}
By the identity (\ref{eq:ident}) 
we have
\begin{equation}                  
	 \int_{S^o} \overline {\theta^o} \Big(\overline {\Delta^1}_{L} \overline{{\phi^1}{'}}
	 \,-\,\overline{{\Delta^1}{'}_{L}}\, \overline {\phi^1} \Big)N_{L}\,dS \nonumber \\
-\, 	 \int_{V^o}  \Big[\overline {\Delta^1}_{L}(\overline {\theta^o}\, \overline {{\phi^1}{'}})_{\!,\,L}
	 \,-\,\overline {\Delta^1}{'}_{L}(\overline {\theta^o}\, \overline {\phi^1})_{,\,L} \Big]\,dV  \,=\,0 \, ,
\end{equation}
and thus
\begin{eqnarray}              
	 \int_{S^o} \overline {\theta^o} \Big(\overline{{\Delta^1}_{L}}\, \overline{{\phi^1}{'}}
	 \,-\,\overline{{\Delta^1}_{L}{'}}\, \overline{\phi^1} \Big)N_{L}\,dS 
\,-\, 	 
	 \int_{V^o} (\overline {\theta^o})_{,\,L} \, \Big(\overline {\Delta^1}_{L} 
	 \overline{{\phi^1}{'}}
	 \,-\,\overline{{\Delta^1}_{L}{'}}\,\overline {\phi^1}\Big)\,dV  \nonumber \\
\,-\, 	 \int_{V^o}  \overline {\theta^o}\, \Big[\overline {\Delta^1}_{L} (\overline{{\phi^1}{'}})_{\!,\,L} 
	 \,-\,\overline{\Delta^1{'}_{\,L}}\,(\overline{{\phi^1}})_{\!,\,L} \Big]\,dV
	  \,=\,0 \, , \end{eqnarray}
\begin{eqnarray}              
	 \int_{S^o} \overline {\theta^o} \Big(\overline{{\Delta^1}_{L}}\, \overline{{\phi^1}{'}}
	 \,-\,\overline{{\Delta^1}{'}_{\!\!\!L}}\, \overline{\phi^1} \Big)N_{L}\,dS 
\,-\, 	 
	 \int_{V^o} (\overline {\theta^o})_{,\,L} \, \Big(\overline {\Delta^1}_{L} 
	 \overline{{\phi^1}{'}}
	 \,-\,\overline{{\Delta^1}{'}_{L}}\,\overline {\phi^1}\Big)\,dV  \nonumber \\
\,+\, 	 \int_{V^o}  \overline {\theta^o}\, \Big(\overline {\Delta^1}_{L} \overline{W^1_{L}{'}}\,-\,\overline{\Delta^1{'}_{\,L}}\,\overline{W^1_{L}} \Big)\,dV
	  \,=\,0 \, .  \end{eqnarray}

Now we substitute the constitutive relation 
$$\overline{{\Delta^1}_{L}}=R_{L N \gamma} 
\overline{u_{\gamma, \, N}} 
- L_{LN} \overline{{\phi^1_{, \, N}}} + \rho_o P_{L} \overline{\theta^1} $$
in the third integral of the last equation.  We obtain

\begin{eqnarray}              
	 \int_{S^o} \overline {\theta^o} \Big(\overline{{\Delta^1}_{L}}\, \overline{{\phi^1}{'}}
	 \,-\,\overline{{\Delta^1}{'}_{\!\!\!L}}\, \overline{\phi^1} \Big)N_{L}\,dS 
\,-\, 	 
	 \int_{V^o} (\overline {\theta^o})_{,\,L} \, \Big(\overline {\Delta^1}_{L} 
	 \overline{{\phi^1}{'}}
	 \,-\,\overline{{\Delta^1}{'}_{L}}\,\overline {\phi^1}\Big)\,dV \,+\, 	 
	 \qquad \qquad \qquad \qquad \qquad   \nonumber \\
\int_{V^o}  \overline {\theta^o}\, 
\Big[\Big(R_{L N \gamma} 
\overline{u_{\gamma, \, N}} 
- L_{LN} \overline{{\phi^1_{, \, N}}} + \rho_o P_{L} \overline{\theta^1}\Big) \overline{W^1_{L}{'}}
- 
\Big(R_{L N \gamma} 
\overline{u{'}_{\!\!\!\gamma, \, N}} 
- L_{LN} \overline{{\phi^1{'}_{\!\!, \, N}}} + \rho_o P_{L} \overline{\theta^1{'}}\Big)
\,\overline{W^1_{L}} \Big] dV
=0 \,.  
\end{eqnarray}
Thus
\begin{eqnarray}  \label{eqnarray:61}              
	 \int_{S^o} \overline {\theta^o} \Big(\overline{{\Delta^1}_{L}}\, \overline{{\phi^1}{'}}
	 \,-\,\overline{{\Delta^1}{'}_{\!\!\!L}}\, \overline{\phi^1} \Big)N_{L}\,dS 
\,-\, 	 
	 \int_{V^o} (\overline {\theta^o})_{\!\!,\,L} \, \Big(\overline {\Delta^1}_{L} 
	 \overline{{\phi^1}{'}}
	 \,-\,\overline{{\Delta^1}{'}_{L}}\,\overline {\phi^1}\Big)\,dV  \nonumber \\
\,+\, 	 \int_{V^o}  \overline {\theta^o}\,\Big[R_{L N \gamma} \Big(
\overline{u_{\gamma, \, N}}\, \overline{{W^1}{'}_{L}}
\,-\,\overline{u{'}_{\!\!\!\gamma, \, N}}
\,\overline{W^1_{L}}\Big)
+ \rho_o P_{L}\Big( \overline{\theta^1}\, \overline{W^1_{L}{'}}- \overline{\theta^1{'}} \, \overline{W^1_{L}}\Big)
\Big] \,dV
	  \,=\,0 .\qquad  \end{eqnarray}
This equation is the analogue of Eq.\cite[(61)]{N:GTT}.

-----------------------------------------------------------------------

Taking the expression for 
\begin{eqnarray}              
\int_{V^o}  \overline {\theta^o}\, 
R_{L N \gamma} \Big( 
\overline{u_{\gamma, \, N}} \, \overline{W^1_{L}{'}}
\,-\,
R_{L N \gamma} 
\overline{u{'}_{\!\!\!\gamma, \, N}}\, \overline{W^1_{L}} \Big)\,dV
  \end{eqnarray}
deduced from (\ref{eqnarray:61}) and inserting this into (\ref{eqnarray:intKLkkky})  yields
\begin{eqnarray}         \label{eqnarray:intKLkkkynew} 
- \int_{V^o} \overline{\theta^o} \Big[\rho_o \Lambda_{L \alpha} \Big(\overline \theta^1 \overline u'_{\alpha, \,L}-\overline \theta^{1'} \overline u_{\alpha, \,L} \Big)\,dV \,=\,\nonumber \\
\,\int_{V^o} \overline{\theta^o} \Big( \overline F_\alpha \overline u{'}_{\alpha}-\overline F{'}_\alpha \overline u_{\alpha} \Big)\,dV \,
+\,\int_{S^o} \overline{\theta^o} \Big( \overline K^1_{L\alpha} \overline u{'}_{\alpha}-\overline K^{1'}_{L\alpha} \overline u_{\alpha} \Big)N_L\,dS \nonumber \\
\,+\,	 \int_{S^o} \overline {\theta^o} \Big(\overline{{\Delta^1}_{L}}\, \overline{{\phi^1}{'}}
	 \,+\,\overline{{\Delta^1}{'}_{\!\!\!L}}\, \overline{\phi^1} \Big)N_{L}\,dS
\,- \, 	 
	 \int_{V^o} (\overline {\theta^o})_{\!\!,\,L} \, \Big(\overline {\Delta^1}_{L} 
	 \overline{{\phi^1}{'}}
	 \,-\,\overline{{\Delta^1}{'}_{L}}\,\overline {\phi^1}\Big)\,dV  \nonumber \\
\,-\, 	 \int_{V^o}  \overline {\theta^o}
 \rho_o P_{L}\Big( \overline{\theta^1}\, \overline{W^1_{L}{'}}- \overline{\theta^1{'}} \, \overline{W^1_{L}}\Big)\,dV
\,-\,
\int_{V^o} (\overline{\theta^o)_{,\,L}} \Big(\overline{K^1_{\!\!L \alpha}}\overline{u'_{\alpha}}
-\overline{K^1{'}_{\!\!L \alpha}} \overline{u_{\alpha}} \Big)  \,dV   \,. \qquad 
\end{eqnarray}
Now inserting (\ref{eqnarray:intKLkkkynew}) in (\ref{eqnarray:577}) yields
\begin{eqnarray}	\label{eqnarray:577new}	
\kappa_{LN\alpha} \int_{V^o} \overline {\theta^o}\Big(\overline {\theta^1}'\overline{\frac{u_{\alpha, \,NL}}{\theta^o}} -
\overline {\theta^1}\overline{\frac{u'_{\alpha, \,NL}}{\theta^o}}\Big)\,dV \,+\,
\kappa^E_{MN} \int_{V^o} \overline {\theta^o}
\Big(\overline {\theta^1{'}}\overline{\frac{\phi^1_{, \, MN}}{\theta^o}} -
\overline {\theta^1}\overline{\frac{\phi^1{'}_{\!, MN}}{\theta^o}}\Big)\,dV \,+\, \nonumber \\
\kappa_{L} \int_{V^o} \overline {\theta^o}
\Big(\overline {\theta^1{'}}\overline{\frac{\theta^1_{, \, L}}{\theta^o}} -
\overline {\theta^1}\overline{\frac{\theta^1_{,\, L}{'}}{\theta^o}}\Big)\,dV \,+\,
\kappa_{MN} \int_{V^o} \overline {\theta^o}
\Big(\overline {\theta^1{'}}\overline{\frac{\theta^1_{, \, NM}}{\theta^o}} -
\overline {\theta^1}\overline{\frac{\theta^1{'}_{\!, NM}}{\theta^o}}\Big)\,dV \,+\, \nonumber \\
+\, p \int_{V^o} \overline \rho_o{\overline{\theta^o}}
\Big( -\overline {\theta^1{'}}P_{M} \overline{W^1{}_{\!\!\!\!M}}
\,+\,
\overline {\theta^1} P_{M} \overline{W^1{'}_{\!\!\!\!M}}\Big) \,dV  \nonumber \\
+\, p \Big[\,\int_{V^o} \overline{\theta^o} \Big( \overline F_\alpha \overline u{'}_{\alpha}-\overline F{'}_\alpha \overline u_{\alpha} \Big)\,dV 
\,+\,
\int_{S^o} \overline{\theta^o} \Big( \overline K^1_{L\alpha} \overline u{'}_{\alpha}-\overline K^{1'}_{L\alpha} \overline u_{\alpha} \Big)N_L\,dS \nonumber \\
\,+\,	 \int_{S^o} \overline {\theta^o} \Big(\overline{{\Delta^1}_{L}}\, \overline{{\phi^1}{'}}
	 \,-\,\overline{{\Delta^1}{'}_{\!\!\!L}}\, \overline{\phi^1} \Big)N_{L}\,dS 
\,-\, 	 
	 \int_{V^o} (\overline {\theta^o})_{\!\!,\,L} \, \Big(\overline {\Delta^1}_{L} 
	 \overline{{\phi^1}{'}}
	 \,-\,\overline{{\Delta^1}{'}_{L}}\,\overline {\phi^1}\Big)\,dV  \nonumber \\
\,-\, 	 \int_{V^o}  \overline {\theta^o}
 \rho_o P_{L}\Big( \overline{\theta^1}\, \overline{W^1_{L}{'}}- \overline{\theta^1{'}} \, \overline{W^1_{L}}\Big)
 \,dV
\,-\,
\int_{V^o} (\overline{\theta^o)_{,\,L}} \Big(\overline{K^1_{\!\!L \alpha}}\overline{u'_{\alpha}}
-\overline{K^1{'}_{\!\!L \alpha}} \overline{u_{\alpha}} \Big)  \,dV\,+\, \nonumber \\
\,+\, 
\int_{V^o} \rho_o {\overline{\theta^o}}  
\Big(\overline {\theta^1}\overline{\frac{\gamma^1{'}}{\theta^o}}
\,-\,\overline {\theta^1{'}} \overline{\frac{\gamma^1}{\theta^o}}
\Big)\,dV \,=\,0 \, .  \qquad \qquad  \qquad          
\end{eqnarray}
Next in the latter equality we transform  the sum of the first four integrals.
Firstly note that by (\ref{eq:LaplTranDEF}) we have
\begin{eqnarray}
	\overline 1\,=\, \int_0^{\infty}\, e^{-pt}\,dt \,=\,1/p \, ,
	\qquad \qquad \qquad \qquad \nonumber \\
	 \quad \theta^o=\theta^o({\bf x})  \; \Rightarrow \; \overline{\theta^o}=\theta^o/p \, , 	\qquad \qquad \qquad \qquad  \nonumber \\
	 \overline{\Big(\frac{h({\bf x},\, t)}{f({\bf x})}\Big)} \, = \, 
	 \frac{1}{f({\bf x})} \int_0^{\infty}\, e^{-pt}h({\bf x},\, t)\,dt \,=\,\frac{1}{f({\bf x})}\,\overline{h({\bf x},\, t)} \,   , \nonumber \\
	 \kappa_{\dots} \int_{V^o} \overline{\theta^o}\Big(\overline{\theta^1{'}}\overline{\frac{f_{\dots}}{\theta^o}} -
\overline{\theta^1}\overline{\frac{f{'}_{\dots}}{\theta^o}}\Big)\,dV \,=\,
	\frac{\kappa_{\dots}}{p} \int_{V^o} \Big(\overline {\theta^1{'}}\,\overline{f_{\dots}} -
\overline{\theta^1}\,\overline{f{'}_{\dots}}\Big)\,dV \, . \nonumber \\
\end{eqnarray}
Hence by these equalities and the constitutive relation for the inìcremental heat flux (\ref{eq:Qflow}), the aforementioned sum of the four integrals equals
\begin{equation}\label{eq:4int}
	\frac{1}{p} \int_{V^o} \Big(\overline {\theta^1{'}}\,\overline{Q^1_{\!L, \,L}} -
\overline{\theta^1}\,\overline{Q^1{'}_{\!\!\!\!L, \,L}}\Big)\,dV \, .
\end{equation}
Again by the identity $$\, ab_{,\,ML} = \big(ab_{,\,M}\big)_{,\,L}-a_{,\,L}b_{,\,M}\,$$
and the divergence theorem, the sum (\ref{eq:4int}) equals
\begin{equation}\label{eq:4int3}
	\frac{1}{p}\Big[ \,\int_{S^o} \Big(\overline {\theta^1{'}}\,\overline{Q^1_{L}} -
\overline{\theta^1}\,\overline{Q^1{'}_{L}}\Big)N_L\,dS \,-\,
\int_{V^o} \Big(\overline {\theta^1{'}_{\!\!\!\!\!, \,L}}\,\overline{Q^1_{\!L}} -
\overline{\theta^1_{\!, \,L}}\,\overline{Q^1{'}_{\!\!, \,L}}\Big)\,dV \,\Big] \,.
\end{equation}
By substituting the sum of the first four integrals in Eq.(\ref{eqnarray:577new}) by (\ref{eq:4int3}), we obtain

\begin{eqnarray}
	\frac{1}{p}\Big[ \,\int_{S^o} \Big(\overline {\theta^1{'}}\,\overline{Q^1_{L}} -
\overline{\theta^1}\,\overline{Q^1{'}_{L}}\Big)N_L\,dS \,-\,
\int_{V^o} \Big(\overline {\theta^1{'}_{\!\!\!\!\!, \,L}}\,\overline{Q^1_{\!L}} -
\overline{\theta^1_{\!, \,L}}\,\overline{Q^1{'}_{\!\!, \,L}}\Big)\,dV \,\Big] 
 \nonumber \\
+\, p P_{M}  \int_{V^o} \overline \rho_o{\overline{\theta^o}}
\Big(-\overline{\theta^1{'}}\overline{W^1{}_{\!\!\!\!M}} 
\,+\,
\overline {\theta^1} \overline{W^1{'}_{\!\!\!\!M}}\Big)\,dV  \nonumber \\
+\, p \Big[\,\int_{V^o} \overline{\theta^o} \Big( \overline F_\alpha \overline u{'}_{\alpha}-\overline F{'}_\alpha \overline u_{\alpha} \Big)\,dV 
\,+\,
\int_{S^o} \overline{\theta^o} \Big( \overline K^1_{L\alpha} \overline u{'}_{\alpha}-\overline K^{1'}_{L\alpha} \overline u_{\alpha} \Big)N_L\,dS \nonumber \\
\,+\,	 \int_{S^o} \overline {\theta^o} \Big(\overline{{\Delta^1}_{L}}\, \overline{{\phi^1}{'}}
	 \,-\,\overline{{\Delta^1}{'}_{\!\!\!L}}\, \overline{\phi^1} \Big)N_{L}\,dS 
\,-\, 	 
	 \int_{V^o} (\overline {\theta^o})_{\!\!,\,L} \, \Big(\overline {\Delta^1}_{L} 
	 \overline{{\phi^1}{'}}
	 \,-\,\overline{{\Delta^1}{'}_{L}}\,\overline {\phi^1}\Big)\,dV  \nonumber \\
\,-\, 	 \int_{V^o}  \overline {\theta^o}
 \rho_o P_{L}\Big( \overline{\theta^1}\, \overline{W^1_{L}{'}}- \overline{\theta^1{'}} \, \overline{W^1_{L}}\Big)
 \,dV
\,-\,
\int_{V^o} (\overline{\theta^o)_{,\,L}} \Big(\overline{K^1_{\!\!L \alpha}}\overline{u'_{\alpha}}
-\overline{K^1{'}_{\!\!L \alpha}} \overline{u_{\alpha}} \Big)  \,dV\,+\, \nonumber \\
\,+\, 
\int_{V^o} \rho_o   
\Big(\overline{\theta^1}\overline{\frac{\gamma^1{'}}{\theta^o}}
\,-\,\overline{\theta^1{'}} \overline{\frac{\gamma^1}{\theta^o}}
\Big)\,dV \,=\,0 \, . \qquad \qquad \qquad        
\end{eqnarray}
The latter is the final form of the {\it theorem of reciprocity of work}, containing all causes and effects.
It generalizes Eq.\cite[(62)]{N:GTT}, and reduces exactly to the latter in case of vanishing initial fields, that is, when the initial configuration is natural.

 

\begin{thebibliography}{Montanaro 99}
 \bibitem{N:RTC} 
  Nowacki, W.  {\it A Reciprocity Theorem for Coupled Mechanical and Thermoelectric Fields in Piezoelectric Crystals}. 
Proc. Vibrations Probl., 6, 1:3-11, 1965. 
\bibitem{N:GTT} 
  Nowacki, W.  {\it Some general Theorems of Thermopiezoelectricity}. 
J. of Thermal Stresses, 1:171-182, 1978. \bibitem{Li:URT} 
  Li, J.Y.  {\it Uniqueness and reciprocity theorems for linear thermo-electro-magnetoelasticity}. 
Q. J. Mech. Appl. Math., 56: 35-43, 2003.
 \bibitem{A:GTTM} 
  Aouadi, M.  {\it The Generalized Theory of Thermo-Magnetoelectroelasticity}. 
Technische Mechanik, 27, 2:133-146, 2007. 
 \bibitem{M:CUF} 
  M\"{u}ller, I.M.  {\it The coldness a universal function in thermoelastic bodies}. 
Arch. Rational Mech. Anal., 41, 319-332, 1971. 
 \bibitem{I:TED} 
  Iesan, D.  {\it Thermopiezoelettricity without energy dissipation}. 
Proc. R. Soc. A, 631:133-656, 2007.  
\bibitem{K:TPT} 
  Kupradze V.D., Gegelia T.G., Basheleishvili M.O. and  Burchuladze T.V. {\it Three-dimensional Problems of the Mathematical Theory of Elasticity and Thermoelasticity}. 
North-Holland, Amsterdam 1979.
\bibitem{C:TRCE} 
  Coleman V.D. and Dill E.H. {\it Thermodynamic restrictions on the constitutive equations of electromagnetic theory}. 
Z. Angew. Math. Phys. Vol. 22, pp. 691-702, 1971.
%
 \bibitem{I:TCS} 
  Amendola, G.  {\it On thermodynamic conditions for the stability of a thermoelectromagnetic system}. 
Math. Meth. Appl. Vol. 23, pp. 17-39, 2000.
%
 \bibitem{I:LST} 
  Amendola, G.  {\it Linear stability for a thermoelectromagnetic material with memory}. 
Math. Mech. Appl. Vol. 59, pp. 67-84, 2001.

\bibitem{MF:ECM} 
  Morro, A. Fabrizio, M.  {\it Electromagnetism of Continuous Media}. 
Oxford University Press. Oxford. 2003.
%
\bibitem{T:NLETE} 
  Tiersten, H.F.  {\it On the Nonlinear Equations of Thermoelectroelasticity}. 
Int. J. Engng Sci. Vol. 9, pp. 587-604. Pergamon Press 1971.

\bibitem{Y:ESF} 
  Yang, J.S.  {\it Equations for Small Fields Superposed on Finite Biasing Fields in a  Thermoelectroelastic Body}. 
IEEE Transactions on Ultrasonics, Ferroelectricts, and Frequency Control, Vol. 50, 187-192, no. 2, February 2003.
%
\end{thebibliography}
\end{document}